# NEOWISE Data Processing and Color Corrections for Near-Earth Asteroid Observations

Samuel A. Myers[1], Ellen S. Howell[1], Yanga R. Fernández[2], Sean E. Marshall[2], Christopher Magri[3], Ronald J. Vervack, Jr.[4], and Mary L. Hinkle[2]
[1] Lunar and Planetary Laboratory, University of Arizona, Tucson, AZ, USA
[2] University of Central Florida, Orlando, FL, USA
[3] University of Maine Farmington, Farmington, ME, USA
[4] Johns Hopkins Applied Physics Laboratory, Laurel, MD, USA

## Abstract

The Wide-field Infrared Survey Explorer and NEOWISE missions are a key source of thermal data for near-Earth asteroids (NEAs). These missions, which utilized a space-based platform in Earth orbit, produced thermal images across four different wavelength bands, W1–W4, with effective wavelengths of 3.4, 4.6, 12, and 22 $\mu$m respectively. Despite its use for NEA observations though, the mission architecture was originally designed to observe stars. Thus, careful data analysis methods are crucial when working with NEA data to account for the differences between these objects. However, detailed information on how to work with these data can be difficult to find for users unfamiliar with the mission. The required information is well documented, but locating it can be challenging, and many details, such as specifics about color corrections, are not fully explained. Therefore, in this work, we provide a set of "lessons learned" for working with NEOWISE data, outline the basics of how to retrieve and process NEOWISE data for NEA investigations, and present an empirical method for color correction determination. We highlight the importance of this process by processing data for three NEAs, finding that nearly half of all available observations should be discarded. Finally, we present simple thermal model results based on different levels of data analysis to highlight how data processing can affect model results.

*Unified Astronomy Thesaurus concepts:* Infrared astronomy (786); Infrared telescopes (794); Near-Earth objects (1092)

## 1. Introduction

The study of near-Earth asteroids (NEAs) is an integral part of planetary science. Understanding NEA orbits, compositions, sizes, and surface properties informs topics such as early solar system evolution, volatile delivery to the early Earth, planetary formation, and, of course, planetary defense (e.g., W. F. Bottke et al. 2002; R. P. Binzel et al. 2015; M. Azadmanesh et al. 2023 and references therein). Much of this information can be gathered through thermal observations of NEAs. Infrared and near-infrared spectra as well as photometric observations in these wavelengths can be interpreted via thermophysical models to probe NEA sizes, albedos, and surface properties (and thus compositions (e.g., M. Delbo et al. 2015; C. Magri et al. 2018).

Today, a key source of NEA thermal data is the Wide-field Infrared Survey Explorer (WISE) spacecraft mission and its follow-on, NEOWISE (E. L. Wright et al. 2010; A. Mainzer et al. 2011, 2014a; NEOWISE Team 2020). WISE was a space-based telescope mission, launched in 2009 December, that measured absolute photometry in four thermal bands. These bands, W1–W4, had effective wavelengths of 3.4, 4.6, 12, and 22 $\mu$m respectively. Observations were made in each band simultaneously, splitting light across the four detectors. The original astrophysics mission scanned the entire sky over a period of roughly 10 months, capturing many NEAs in the process. At the end of this time, depletion of the cryogenic coolant degraded the sensitivity of the longer wavelength detectors. This eventually led the mission to take data for a brief time in a 3-band configuration, using only bands W1–W3. Soon after, full depletion of the coolant resulted in the mission entering its "post-cryo" phase, where only bands W1 and W2 were operable. Together, these three mission phases spanned 2010 January to 2011 February. In 2013 December, the spacecraft was reactivated as NEOWISE, taking data in the NEOWISE-R mission phase using only bands W1 and W2, with an explicit focus on NEA detection and characterization. This mission phase lasted until the depletion of propellant led to the spacecraft reentering Earth's atmosphere in 2024 September. Note that in this paper we will use "NEOWISE" as a shorthand to refer to the entire mission including all mission phases.

Altogether, the mission gathered NEA data over a roughly 12 yr period spanning four mission phases: full four-band WISE, 3-band WISE, two-band post-cryo, and two-band NEOWISE-R. NEOWISE data sets thus provide invaluable insights into the NEA population, and have been widely utilized for NEA investigations. However, accurate use of these data requires careful data analysis methods. For one, the W1 and W2 bands both contain both thermal as well as reflected light, and the W1 band in particular has lower sensitivity, complicating photometric measurements. However, the primary concern is that the mission architecture was originally designed for observations of stars rather than NEAs. As a result, the mission architecture is optimized for observations of stars, which can be considered single-temperature point sources. This is in contrast to NEAs, which have a varied surface temperature distribution. Therefore, these two object types have vastly different temperatures and thus different

Table 1
Properties of NEAs Used as Examples in This Work

| Object | Spectral Type | Shape | Rotation (hr) | Diameter (km) | H-magnitude | Visible Albedo |
|---|---|---|---|---|---|---|
| (53319) 1999 JM8 | P (Tholen) | Irregular | 168 (NPA) | ∼7 | ∼15–16.5 | 0.01–0.06 |
| (85989) 1999 JD6 | L/K (Bus-DeMeo) | Contact Binary | 7.67 | ∼0.6–3.0 | ∼16.6–17.1 | 0.01–0.31 |
| (137032) 1998 UO1 | S (Bus-DeMeo) | Spheroidal | 2.9 | ∼1.1 | ∼16.5 | 0.15–0.50 |

**Note.** Spectral types are given in either the Tholen (D. J. Tholen & M. A. Barucci 1989) or Bus-DeMeo (F. E. DeMeo et al. 2009) taxonomies. NPA is nonprincipal axis rotator. Values for JM8 are taken from L. A. Benner et al. (2002), V. Reddy et al. (2012), R. P. Binzel et al. (2019), J. R. Masiero et al. (2020), and S. A. Myers et al. (2024). Values for JD6 are taken from R. P. Binzel et al. (2001), E. S. Howell et al. (2008), D. Polishook & N. Brosch (2008), H. Campins et al. (2009), C. Thomas et al. (2011), V. Reddy et al. (2012), A. Mainzer et al. (2014b), C. Nugent et al. (2016), S. E. Marshall et al. (2017), R. P. Binzel et al. (2019), D. Kuroda et al. (2021), and S. A. Myers et al. (2024). Values for UO1 are taken from S. D. Wolters et al. (2008), C. Thomas et al. (2011), and S. A. Myers et al. (2024).

spectral energy distributions. Together, these factors make NEA observations more difficult to interpret.

The required information for proper NEA data analysis is well documented, but it is spread throughout multiple sources, and some key details are missing. Therefore, in this work, we set out to provide an easily consolidated roadmap of key points for using NEOWISE data for NEA investigations. Specifically, in this paper, we examine three different NEAs to highlight the required steps for NEOWISE data analysis. (Note that all information provided here, with the exception of the discussion on color corrections (Section 3), would also be applicable to main-belt asteroids (MBAs).) These three NEAs represent a range of asteroid spectral types and physical characteristics and were chosen because they are well characterized. Previous works have presented ground-based spectral data, radar data, and many modeling results for each object. Thus, these objects are well understood with well constrained properties and are well suited as examples for comparing with available NEOWISE data.

The NEA (53319) 1999 JM8 is a large, dark, irregularly shaped, nonprincipal axis rotator (NPA). The NEA (85989) 1999 JD6 is a medium brightness, contact binary. The NEA (137032) 1998 UO1 is spheroidal and rapidly rotating. Object spectral types, rotation rates, diameters, H-magnitudes, and visible albedo ranges are listed in Table 1. Throughout this work, we will refer to each object by its alphanumeric designation (i.e., JM8).

For each of these objects, we step through the required data processing steps, highlighting how the data change at each point in the process. The data are also analyzed using a simple thermal model to show how model results can be affected by changes in the data processing procedure. In Section 2, we walk through the data processing procedure in detail. In Section 3, we discuss an empirical method for selecting color correction values for the NEOWISE data. In Section 4, we walk through how simple thermal model results based on the NEOWISE data change at each step of the data processing procedure. Finally, we highlight our conclusions in Section 5.

## 2. Data Processing Procedure

Accurately working with data from the NEOWISE mission requires using a careful data-vetting process. This ensures that only noncontaminated observations of the actual object in question are used, that multiple observations are averaged together appropriately, and that the proper corrections are made to adjust for the effects of the filter bandpasses. This last step is especially important as the mission architecture was originally designed to look at stars, which can be thought of as single-temperature point sources, as opposed to asteroids, which are more accurately described by a surface with a temperature distribution. These differing spectral energy distributions therefore require careful correction when using NEOWISE photometry for NEAs. All of the required information for properly working with the data can be found in the mission handbooks (E. L. Wright et al. 2010) and in the mission's supporting literature (A. Mainzer et al. 2011, 2014a). However, this information is spread across multiple sources and can be difficult to parse for users unfamiliar with the mission.

Here, we walk through the key steps required to process NEOWISE NEA observations. Note that all steps listed below are also applicable to MBA observations. Many of these steps are described in the mission literature; however, we also highlight some additional checks that we found to be beneficial when working with NEOWISE data. Together, these steps ensure only noncontaminated data are used for analysis. These steps are as follows.

1. *Finding actual observations.* The first step is to ensure that only observations of the actual object in question are used. This requires pulling reported observations from the Minor Planet Center (MPC)[5] for comparison with the NEOWISE database. This will ensure that only observations where the object was present and centered within the instrument's field of view will be returned.
   (a) Pull observations from the MPC for the object in question, filtered with the WISE observatory code of C51.
   (b) Save these observations, and convert them into the IPAC format.[6]
   (c) In the Infrared Science Archive (NEOWISE Team 2020), locate the L1b Single Source Catalogs for the NEOWISE mission. There are four catalogs, each corresponding to the different mission phases: WISE All-Sky, WISE 3-Band Cryo, WISE Post-Cryo, and NEOWISE-R.
   (d) Use the Multi Object Search function, and upload the IPAC formatted list of observations from the MPC. Use a conical search radius of 0″.3, and enter the additional constraint of `abs(mjd-mjd_01) < 2./86400`. This will ensure that only observations from the NEOWISE database that are within 0″.3 on the sky and within 2 s in time of the reported MPC observations will be returned.[7] The "One To One Match" option does not need to be selected since the catalog is already only being searched where the object is known to be.

---
[5] https://www.minorplanetcenter.net/db_search
[6] https://irsa.ipac.caltech.edu/applications/DDGEN/Doc/ipac_tbl.html
[7] https://wise2.ipac.caltech.edu/docs/release/allsky/expsup/sec2_4e.html

2. *Checking data flags.* The NEOWISE pipeline applies automatic flags to some observations that may be contaminated or are of overall low quality. The exact flags and their format vary across the different mission phases, but each is important for ensuring only noncontaminated observations are used.[8] The flags are as follows.
   (a) `cc_flags`. This flag identifies any potential contamination within the image. WISE data have a single `cc_flags` value while other mission phases have an individual value listed for each band. Any observation with a value other than `0` in any band should be discarded for all bands. Common values are `d` indicating a diffraction spike, `h` indicating a halo, or `g` indicating optical glint.
   (b) `ph_qual`. This flag serves to classify the flux signal-to-noise ratio (SNR) of the source detected in each given band. `A` indicates SNR greater than 10, `B` indicates SNR between 3 and 10, and `C` indicates SNR between 2 and 3. `U` indicates the upper magnitude limit was reached, and `X` indicates that a measurement of the source was not possible in that band. Any observation with a value of `U` or `X` in any band should be discarded for all bands. Note that for mission phases with missing bands (i.e., NEOWISE-R where only W1 and W2 were active) the missing bands will be listed as `X`. In these cases, only the active bands need to be considered (i.e NEOWISE-R observations with an `X` value in the W3 and W4 band can be kept).
   (c) `qual_frame`. This flag is an overall quality score that is assigned to each image. Note that only four-band WISE and NEOWISE-R data have a `qual_frame` value. (Furthermore, for four-band WISE data, this column is not selected by default in the Multi Object Search). This value is an integer ranging from 0 to 10, with 0 being the lowest score and 10 being the highest. However, in practice, this value is almost always 0, 5, or 10. Any observation with a value of `0` should be discarded.
3. *Checking the raw images.* In addition to the flags assigned automatically by the pipeline, images should also be checked manually by visual inspection. The pipeline is not able to catch all contaminated images, and thus, this step is essential to filter out all spurious and contaminated detections. Images can be checked by searching for the object in the WISE Image Service,[9] entering the object of interest, and selecting the desired mission phases. Images should be examined primarily to check that the object is not overlapping with another object in the frame. However, in some instances, diffraction spikes, missing objects, and streaked objects may be identified (Figure 1). All images contaminated in these ways should be discarded. Note that when working with population level data, the errors introduced from a failure to inspect the images manually are likely to cancel out in the aggregate. This step is primarily a concern when considering individual objects. However, when working with larger data quantities, three additional flags can be checked to further filter results. However, these checks are not robust. Some spurious or contaminated detections will not be flagged, and some nonspurious detections will be flagged. Therefore, careful manual inspection is still required. These additional flags are as follows.
   (a) `nb`: This flag is an integer that indicates the number of point-spread functions identified by the pipeline. Any value greater than 1 may be a sign of overlapping objects.
   (b) `rchi2`: These are a set of flags associated with every band present. A null value may indicate a nondetection.
   (c) `qual_frame`: Same as above. Some missing objects and diffraction spikes may have a value greater than 0 but less than 10. Therefore, observations with values between 0 and 10 (usually a value of 5) may be good candidates for further visual inspection.
4. *Data averaging.* Depending on the number of remaining data points, it will likely be beneficial to average together observations for analysis. In most cases, the time span of NEOWISE observations will be longer than the object's rotation rate, and thus, rotational phase effects will be averaged out. Therefore, it is best to average together close-in-time observations, utilizing natural gaps in the data when possible. When averaging, it is important to keep in mind that NEOWISE data uncertainties are dominated by systematic errors and not statistical noise. Therefore, for each band, the weighted average should be used, adopting as the new $1\sigma$ data uncertainties the variance of the overall data set divided by the square root of the number of observations minus 1. Furthermore, as an additional data check, an additional requirement can be imposed that each data set requires at least three observations in a given band with an uncertainty (the `sigmpro` value in the database) of $\sigma < 0.25$ (J. R. Masiero et al. 2020). This ensures any potentially remaining spurious detections are averaged out, but single observations of bright objects with high SNR can be considered robust if they pass the checks laid out above.

We apply the above procedure to our three example objects to highlight the importance of following these steps. We start by simply pulling all data returned for each object through a basic Moving Object Search of the L1b Single Source Catalogs for each mission phase (NEOWISE Team 2020). We then apply the above steps, filtering out spurious and contaminated data at each step. The results of this procedure are shown in Table 2. Overall, we find that roughly half of all observations are filtered out through the above process. In particular, 25% of the observations are filtered out by properly checking the MPC, an additional ∼17% are filtered out by vetting the flags, and ∼7% are filtered out by manual inspection of the images. Thus, following these data processing steps is absolutely critical for properly filtering NEOWISE observations in order to avoid incorporating false or contaminated detections in data analysis.

## 3. Color Corrections

All NEOWISE data require a color correction value when converting between NEOWISE magnitudes and absolute flux

---

[8] https://wise2.ipac.caltech.edu/docs/release/neowise/expsup/sec2_1a.html
[9] https://irsa.ipac.caltech.edu/applications/wise/?__action=layout.showDropDown&

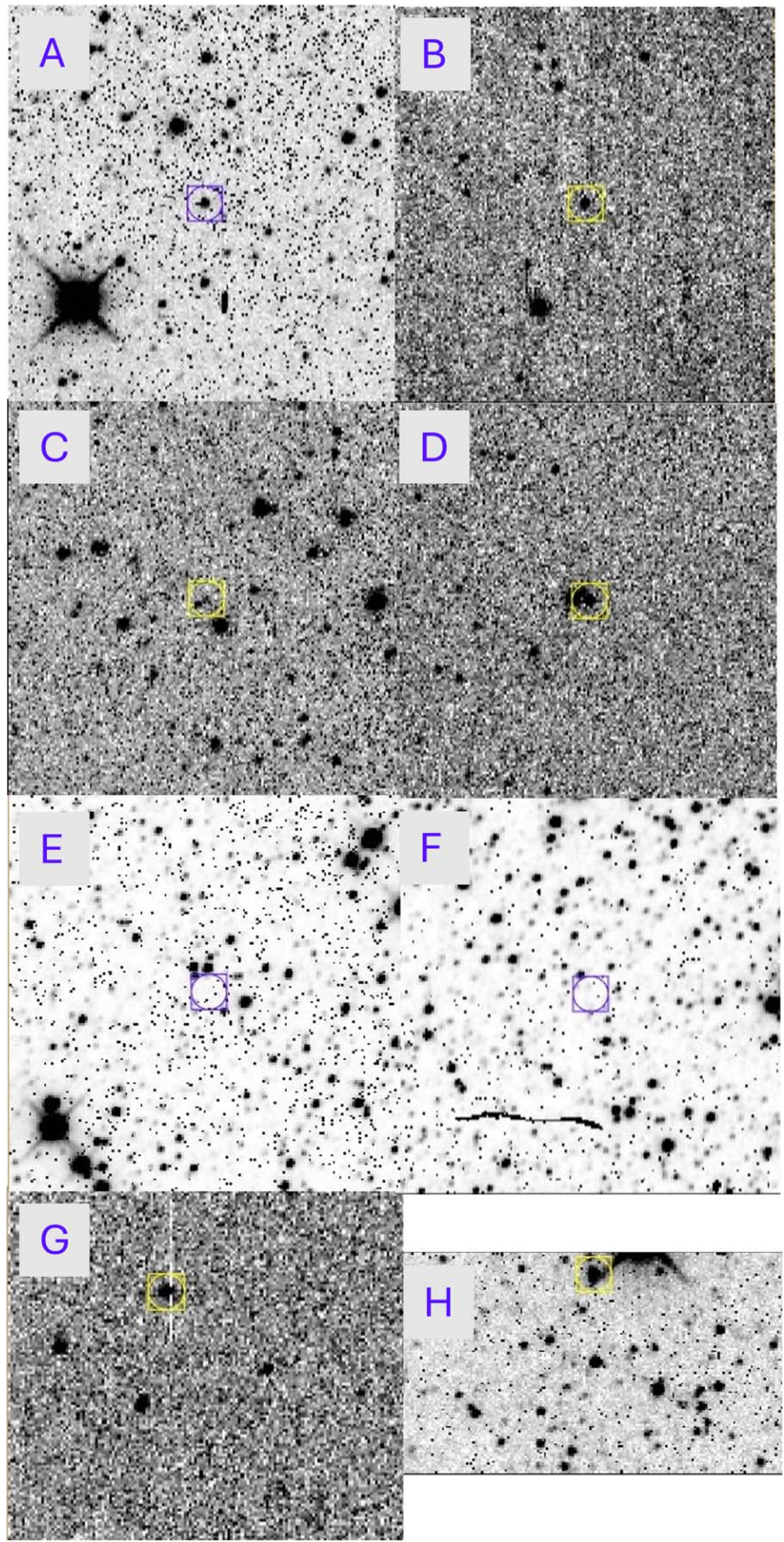

**Figure 1.** Raw NEOWISE-R images. The reticle shows the location from which the automatic pipeline extracts a magnitude value. (A) and (B): UO1 W2 and JM8 W2. Example of a nonspurious detection with the object clearly visible. (C) and (D): JD6 W2. Examples of contamination by a background source. (E) and (F): JD6 W1. Examples of a spurious detection with no object visible. (G): JD6 W2. Example of a streaked image, in this case likely due to a bad column. (H): JM8 W2. Example of a diffraction spike overlapping the primary object. All frames shown here were *not flagged* by the automatic pipeline.

Table 2
The Number of Observations that Pass through Each Step of the Data Processing Procedure

| Object | Phase | No. of Initial | No. of MPC | No. of Flags | No. of Images | Total (%) |
|---|---|---|---|---|---|---|
| JM8 | WISE | 3 (1) | 0 (0) | 0 (0) | 0 (0) | 0 |
| JM8 | Post-Cryo | 1 (1) | 0 (0) | 0 (0) | 0 (0) | 0 |
| JM8 | NEOWISE-R | 131 (20) | 104 (13) | 96 (12) | 82 (12) | 63 |
| JM8 | Total | 135 (22) | 104 (13) | 96 (12) | 82 (12) | 61 |
| JD6 | WISE | 24 (5) | 24 (5) | 11 (4) | 11 (4) | 46 |
| JD6 | NEOWISE-R | 115 (13) | 91 (14) | 61 (12) | 54 (10) | 47 |
| JD6 | Total | 139 (18) | 115 (19) | 72 (16) | 65 (14) | 47 |
| UO1 | WISE | 1 (1) | 0 (0) | 0 (0) | 0 (0) | 0 |
| UO1 | Post-Cryo | 7 (1) | 2 (1) | 1 (1) | 1 (1) | 14 |
| UO1 | NEOWISE-R | 18 (4) | 4 (2) | 4 (2) | 4 (2) | 22 |
| UO1 | Total | 26 (6) | 6 (3) | 5 (3) | 5 (3) | 19 |

**Note.** The number of averaged data sets produced at each stage is given in parenthesis. Object gives the name of the object, and phase denotes which mission phase the observations are taken from, with total being the sum of all mission phases. Initial is the initial number of observations pulled from the moving object search. MPC is the number of observations that remain after checking against the MPC (Step 1). Flags is the number of observations that remain after checking the data flags (Step 2). Images is the number of observations that remain after checking the raw images (Step 3). % Total is the percentage of initial observations that remains after all three vetting steps.

Table 3
Table of NEOWISE Color Corrections for Corresponding Blackbody Temperatures

| $F_\nu$ | W1 | W2 | W3 | W4 |
|---|---|---|---|---|
| $B_\nu(100)$ | 17.2062 | 3.9096 | 2.6588 | 1.0032 |
| $B_\nu(141)$ | 4.0882 | 1.9739 | 1.4002 | 0.9852 |
| $B_\nu(200)$ | 2.0577 | 1.3448 | 1.0006 | 0.9833 |
| $B_\nu(283)$ | 1.3917 | 1.1124 | 0.8791 | 0.9865 |
| $B_\nu(400)$ | 1.1316 | 1.0229 | 0.8622 | 0.9903 |
| $B_\nu(566)$ | 1.0263 | 0.9919 | 0.8833 | 0.9935 |
| $B_\nu(800)$ | 0.9884 | 0.9853 | 0.9125 | 0.9958 |

**Note.** Adapted from Table 1 in E. L. Wright et al. (2010). Left column is blackbody temperature in kelvins. Remaining columns are corresponding color corrections ($f_c$) for bands W1 − W4.

units. This conversion is given by the equation[10]

$$F_\nu[\text{Jy}] = \left(\frac{F_{\nu 0}}{f_c}\right) 10^{-m/2.5} \quad (1)$$

where $F_\nu[\text{Jy}]$ is the final flux in Jansky units, $F_{\nu 0}$ is the set of zero-magnitude flux densities given in Table 1 of E. L. Wright et al. (2010), $f_c$ is the color correction value, and $m$ is the NEOWISE magnitude. The color correction value is essentially a spectral slope correction for the bandpass filters, correcting for instrumental sensitivity effects. The color correction is based on a convolution of the input spectrum and the bandpass filters and is derived based on the source's blackbody temperature. Note that red sources, defined as sources with a steeply rising mid-infrared spectrum, which thus includes most NEAs, require an additional correction factor of 0.90 to $F_\nu$ for the W4 band (E. L. Wright et al. 2010).

E. L. Wright et al. (2010) give a list of color correction values corresponding to various blackbody temperatures (Table 3). However, no guidance is given on how to calculate the appropriate effective blackbody temperature for a given source. This is particularly relevant as the values given are calculated for stars, which are point-source blackbodies, but NEAs instead have surface temperature distributions. Therefore, because these two object types have vastly different spectral energy distributions, selecting the correct effective blackbody temperature for the source object is critical for calculating the correct color correction value.

Various methods for selecting these values are used throughout the literature. Common methods include incorporating the color correction step into the thermal models used for analysis directly (e.g., J. R. Masiero et al. 2021), using a simple thermal model to calculate an average surface temperature (e.g., M. Varakian et al. 2024), or simply picking a single representative value for all sources. Still, many other works do not explain how or if their color corrections are chosen and applied. However, each of these methods comes with certain drawbacks. Picking a single value, while simple, is an oversimplification that is likely to produce incorrect results for many objects. This is not as problematic when working with population averages, but introduces significant limitations for analyzing single objects. Using a simple thermal model to first calculate an average surface temperature introduces the potential for circularity. In this case, the data are used to produce model results that are used to adjust the data that are used to provide further modeling results. Furthermore, the simplifications inherent in these models may not adequately represent the asteroid's surface, thus leading to the use of an incorrect spectral energy distribution.

The most robust method for applying color corrections is to apply them to modeled fluxes rather than to the data directly. However, this may require significant alterations to the model code for proper implementation, and thus may not be practical for all users. This is especially true for those using simple thermal models where computational speed and ease of use are key considerations. Furthermore, this may be a limitation for those simply wishing to use NEOWISE data alongside other thermal or lightcurve data in more complex thermophyscial models. In these cases, alterations to the model code may be impractical or interfere with the processing of other data sets that do not require such corrections.

In the interest of avoiding the issues with these methods, and maintaining ease of use of the NEOWISE data, we have developed an empirical technique for determining the color corrections for NEAs. Our technique incorporates temperature uncertainty into the data uncertainty. Although less robust than directly applying the color corrections to modeled flux values, this method allows for an accurate determination of the color correction temperature for each source, avoids adding more complication to models, and avoids model circularity, at the expense of introducing more uncertainty into model results. For our technique, we start with a theoretical blackbody temperature in the relevant bandpasses and then add additional uncertainty based on the average error between an asteroid's

---

[10] https://wise2.ipac.caltech.edu/docs/release/allsky/expsup/sec4_4h.html

**Table 4**
Table of SpeX Effective Blackbody Temperature Fits and Corresponding Calculated Theoretical Blackbody Temperatures, Sorted by Heliocentric Distance

| Object | SpeX Fit Temperature (K) | Calculated Temperature (K) | Temperature Difference (K) | $r_H$ (au) | Phase Angle (deg) |
|---|---|---|---|---|---|
| SD220 | 258 | 272 | −14 | 0.9909 | 82.3 |
| UO1 | 268 | 268 | 0 | 0.9963 | 90.2 |
| SD220 | 273 | 270 | 3 | 1.0017 | 79.6 |
| FG3 | 270 | 274 | −4 | 1.0320 | 61.4 |
| JD6 | 257 | 271 | −14 | 1.0330 | 73.2 |
| JD6 | 269 | 269 | 0 | 1.0398 | 76.0 |
| FG3 | 250 | 272 | −22 | 1.0476 | 58.6 |
| UO1 | 260 | 263 | −3 | 1.0775 | 48.6 |
| UO1 | 265 | 260 | 5 | 1.0779 | 48.5 |
| UO1 | 268 | 260 | 8 | 1.0780 | 48.4 |
| UO1 | 260 | 260 | 0 | 1.0781 | 48.4 |
| UO1 | 262 | 260 | 2 | 1.0782 | 48.4 |
| FG3 | 270 | 268 | 2 | 1.0783 | 35.9 |
| UO1 | 275 | 260 | 15 | 1.0783 | 48.4 |
| FG3 | 265 | 265 | 0 | 1.0990 | 28.2 |
| Ivar | 262 | 254 | 8 | 1.1387 | 60.4 |
| FG3 | 260 | 260 | 0 | 1.1422 | 14.0 |
| UO1 | 252 | 252 | 0 | 1.1500 | 46.4 |
| UO1 | 252 | 252 | 0 | 1.1505 | 46.4 |
| FG3 | 260 | 257 | 3 | 1.1742 | 7.2 |
| Eros | 260 | 245 | 15 | 1.1883 | 51.2 |
| FG3 | 248 | 251 | −3 | 1.2256 | 19.2 |
| Eros | 258 | 241 | 17 | 1.2304 | 31.8 |
| Ivar | 257 | 236 | 21 | 1.3214 | 30.8 |
| Ganymed | 249 | 235 | 14 | 1.3236 | 48.9 |
| Ivar | 233 | 233 | 0 | 1.3589 | 26.5 |
| Ivar | 228 | 228 | 0 | 1.4092 | 42.3 |
| Eros | 236 | 217 | 19 | 1.5242 | 39.0 |
| Ivar | 245 | 215 | 30 | 1.5893 | 20.2 |
| Eros | 225 | 200 | 25 | 1.7803 | 30.6 |

**Note.** Object is the object observed with SpeX. SpeX fit temperature is the effective blackbody temperature fit to the SpeX spectrum. Calculated temperature is the theoretical blackbody temperature calculated using Equation (2). Temperature difference is the SpeX fit temperature minus the calculated temperature. $r_H$ is the heliocentric distance of the SpeX observation in astronomical units, and phase angle is the solar phase angle of the SpeX observation in degrees. All observations listed here were chosen solely based on their observing geometries, which closely match trends in NEOWISE observation geometries.

real surface temperature distribution and its theoretical blackbody temperature. To quantify this error, we compared blackbody curves of various temperatures to spectral data of real objects. Because the color correction is an adjustment based on the shape of the input spectrum, this allows us to quantify the temperature difference between a real object's spectrum and a theoretical blackbody spectrum.

This was done by calculating blackbody curves and comparing them to NEA spectra obtained with the SpeX instrument on the NASA Infrared Telescope Facility (IRTF; J. Rayner et al. 2003). All SpeX data were obtained as part of an ongoing project to analyze the use of thermophysical models for characterizing NEAs. For each object observed, we collect multiple data sets across multiple nights of observation. Data were collected using both the prism and long-wavelength cross-dispersed observing modes. Thus, we collected data from 0.8 to 4.2 $\mu$m before the SpeX upgrade in 2014, and from 0.7 to 5.3 $\mu$m after the SpeX upgrade. In addition to the object, we observed a nearby G-type dwarf star and a well-characterized solar analog for data calibration. All observations were processed using the Spextool software package (M. C. Cushing et al. 2004) to apply flat-field corrections, sum images, and extract spectra. Finally, all data were corrected for terrestrial atmospheric absorption (S. D. Lord 1992). Further details on the data processing procedure used can be found in E. S. Howell et al. (2018) and S. A. Myers et al. (2024).

We select SpeX spectra to use for our comparisons that have viewing geometries similar to those most often found in the NEOWISE data set. NEOWISE data are often highly correlated between the heliocentric distance of observation and the solar phase angle, due to the mission's observing geometry. As a result, most NEOWISE data are taken at high solar phase angles and low heliocentric distances, or vice versa. Objects near 1 au have higher solar phase angles, $\gtrsim$40°, while objects closer to 2 au have lower solar phase angles. A subset of SpeX observations are selected that mimic this observing geometry. These include observations of the NEAs (433) Eros, (1036) Ganymed, (1627) Ivar, (163899) 2003 SD220, and (175706) 1996 FG3, as well as UO1 and JD6 (Table 4). For each observation used, we calculate a theoretical blackbody temperature according to the equation

$$\sigma_{sb} T^4 = \frac{L_\odot (1-A)}{16 \pi r_H^2} \qquad (2)$$

where $\sigma_{sb}$ is the Stefan-Boltzmann constant, $L_\odot$ is the solar luminosity, $A$ is the Bond albedo, and $r_H$ is the heliocentric distance. Here, unit emissivity is assumed. The Bond albedo is

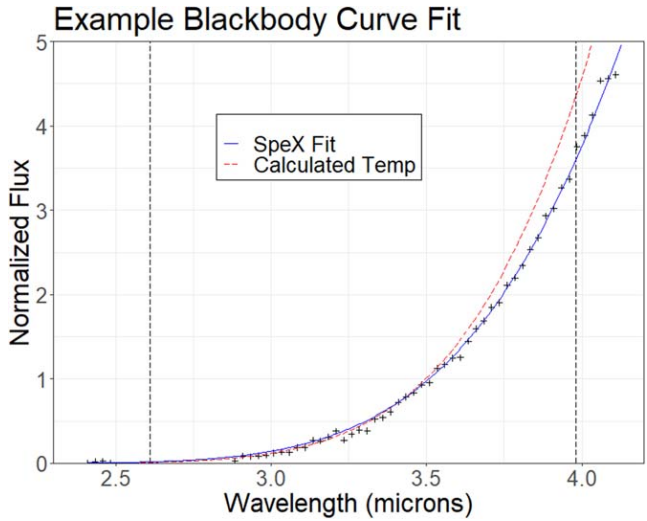

**Figure 2.** Example of an effective blackbody curve fit to a SpeX data set. The data are an observation of Ivar taken on 2013 September 15. The solid blue line is the effective blackbody curve fit to the SpeX data with a temperature of 257 K. The dashed red line is the theoretical blackbody curve calculated using Equation (2) with a temperature of 236 K. The inputs used are $r_H = 1.3214$ au, $p = 0.15$, and $G = 0.60$. The vertical dashed lines show the outer limits of the W1 bandpass filter. The x-axis is wavelength in microns. The y-axis is normalized flux, normalized at 3.4 $\mu$m.

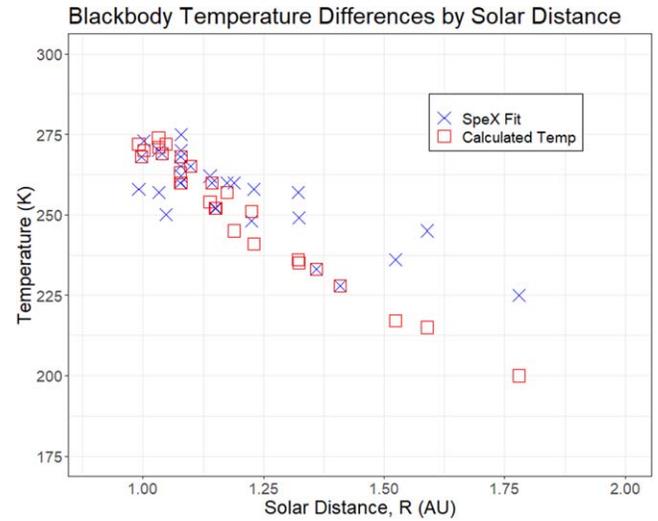

**Figure 3.** Figure of SpeX effective blackbody temperature fits and corresponding calculated theoretical blackbody temperatures. Each vertical line of symbols is one SpeX observation, sorted by heliocentric distance. The cross is the effective blackbody temperature found by fitting blackbody curves to the SpeX data and the box is the theoretical blackbody temperature calculated using Equation (2). The difference between the two never exceeds 30 K. For $r_H < 1.25$ au, the direction of offset is random; for $r_H > 1.25$ au, the calculated theoretical value is always less than the fitted effective value. All observations shown here were chosen solely based on their observing geometries, which closely match trends in NEOWISE observation geometries.

estimated using the method described in L. A. Lebofsky & J. R. Spencer (1989):

$$A = (0.29 + 0.684G)p \quad (3)$$

where $p$ is the visual geometric albedo, and $G$ is the slope parameter in the HG magnitude system (E. Bowell et al. 1989).

We note that Equation (2) obviously omits certain key considerations, such as solar phase angle and emissivity. However, as shown below, in all cases, this equation yields temperatures within 30 K of the actual temperatures observed in SpeX spectra. This is likely the result of the high correlation in the NEOWISE data between solar distance and solar phase angle, resulting in the canceling of errors. Regardless, the procedure laid out here meets our criteria of providing a simple and easily applicable method for approximating the color correction factor and its associated uncertainties, including those implicit within Equation (2).

The SpeX spectra are then compared to these blackbody curves across the bandpass of the W1 filter, where the color correction effects are the strongest for NEA data. We then fit subsequent blackbody curves to the SpeX data to identify the average error between the calculated theoretical blackbody temperature and the best-fitting effective blackbody curve. We chose W1 as the comparison filter because it is the most sensitive to the color correction value. This is because of the band's low sensitivity and the high percentage of reflected light often found in this bandpass. As a result, the W1 bandpass requires much larger color correction values than the other bands (Table 3) and is thus much more sensitive to changes in the chosen blackbody temperature. An example of this process is shown in Figure 2.

By applying this procedure to all observations listed in Table 4, we find that the absolute value of the temperature difference between the calculated theoretical blackbody temperature and the effective blackbody temperature fit to the SpeX spectra never exceeds 30 K. Furthermore, we see that for observations taken at heliocentric distances greater than ~1.25 au the difference between the effective blackbody temperature fit to the SpeX data and the calculated theoretical blackbody temperatures is always positive. That is, the effective blackbody fit temperature is higher than the calculated theoretical blackbody temperature. Therefore, we find that the error between an asteroid's real surface temperature distribution and its theoretical blackbody temperature is no more than 30 K. For observations taken at heliocentric distances less than 1.25 au, this uncertainty interval is centered on the calculated theoretical temperature, and for observations taken at heliocentric distances greater than 1.25 au, this uncertainty interval extends above the calculated theoretical temperature. All of these results are shown in Table 4 and Figure 3. Finally, we note that our data only extend up to heliocentric distances of roughly 2 au, and thus, this method may not be robust past this point, including for most MBAs. However, given that most usable NEOWISE NEA data are taken at heliocentric distances of 2 au or smaller, including all the data highlighted in this work, we do not consider this a significant limitation for the NEA population. The treatment here could be extended to identify similar differences ranges for MBA data. These data would not be affected by the same high solar phase angle effects as NEA data. However, this extension of the color correction procedure to larger solar distances is beyond the scope of this work.

Thus, our new method for applying color corrections is as follows.

1. Calculate the theoretical blackbody temperature for the averaged NEOWISE observation according to Equation (2).
2. Apply a temperature uncertainty range of 30 K. For observations with heliocentric distances less than 1.25 au, this 30 K range should be centered on the calculated theoretical temperature. For observations with heliocentric distances greater than 1.25 au, this 30 K range should be added to the calculated theoretical temperature. Note that if non-NEOWISE spectral data in the W1 and W2 bandpass range are available for the object of interest, the temperature may be computed directly by comparing blackbody curves to this spectrum of the object, as long as the viewing geometry is similar to that of the NEOWISE observations.

3. Calculate color correction factors corresponding to the calculated temperature range. This is done by convolving the corresponding blackbody spectra with the bandpass filters. Contact the authors of this manuscript for the code for the implementation of this step.

4. Apply these color correction factors during the conversion process described by Equation (1). For W4 observations, incorporate the additional red slope correction (E. L. Wright et al. 2010). This will produce a set of data points with overlapping uncertainties that are added to obtain the final flux values and associated uncertainties.

### 4. Data Processing Effects on Model Results

In this section, we highlight the effects these data processing steps can have on model analysis and interpretation for various objects. Here, we take three different NEAs (JM8, JD6, and UO1) and analyze them using a simple thermal model at each step of the data analysis process. Thus, we analyze all data available from a simple Moving Object Search, all data remaining after checking against the MPC, all data remaining after checking the data flags, and all data remaining after checking the images. For all these steps, we apply a color correction corresponding to a temperature of 200 K. This is chosen as a broadly "representative" NEA temperature value for which color correction values are provided directly by E. L. Wright et al. (2010; Table 3). Finally, we apply our color correction procedure to the remaining data and analyze them as well.

#### 4.1. Simple Thermal Model

The exact model we use for this work is based on the Standard Thermal Model (L. A. Lebofsky et al. 1986; L. A. Lebofsky & J. R. Spencer 1989; J. R. Spencer et al. 1989) and the Near-Earth Asteroid Thermophysical Model (NEATM) (A. W. Harris 1998). The model assumes a spherical shape for the object as well as an equatorial subsolar and subobserver point. The model also assumes a constant, prograde rotation period that allows for a rudimentary incorporation of thermal inertia. Fixed model inputs include the object's rotation period, its H-magnitude, emissivity, Earth–object distance, Sun–object distance, solar phase angle, and a visible-to-near-IR reflectance ratio that scales between visible and infrared albedo. Literature values are used for the rotation period and H-magnitude. The emissivity is held constant at 0.90 for all objects. Ephemerides are calculated for each observation using ESA NEODyS-2.[11] The visible-to-near-IR reflectance ratio is calculated using our SpeX prism data and visible object data.

The model itself is a forward model, producing model thermal spectra for a given set of input parameters. We run models over a range of three free-floating input parameters: the visible geometric albedo, the thermal inertia, and the beaming parameter. The visible geometric albedo is given as a dimensionless value from 0.0 to 1.0. The thermal inertia is a positive value given in SI units of $J\,m^{-2}\,s^{-1/2}\,K^{-1}$ (referred to here as thermal inertia units or TIU). The beaming parameter is a dimensionless scaling factor that gives a first-order accounting of the various model assumptions. The beaming parameter thus scales the final model flux to attempt to account for effects that are not modeled explicitly, such as self-shadowing, deviations from a spherical shape, surface roughness, and nonzero obliquities. The beaming parameter is generally a value between 0.5 and 2.5. Because beaming parameter is expected to change across observations, we run models across a range of physically plausible beaming parameters that produce well-fitting spectra. Therefore, for each observation, we seek to identify the ranges of visible albedos and thermal inertias that produce models that fall within the $1\sigma$ uncertainties of the NEOWISE data (Figure 4). For more details on the model and model fitting procedure employed here, see E. S. Howell et al. (2018) and S. A. Myers et al. (2024).

---
[11] https://newton.spacedys.com/neodys/index.php?pc=3.0

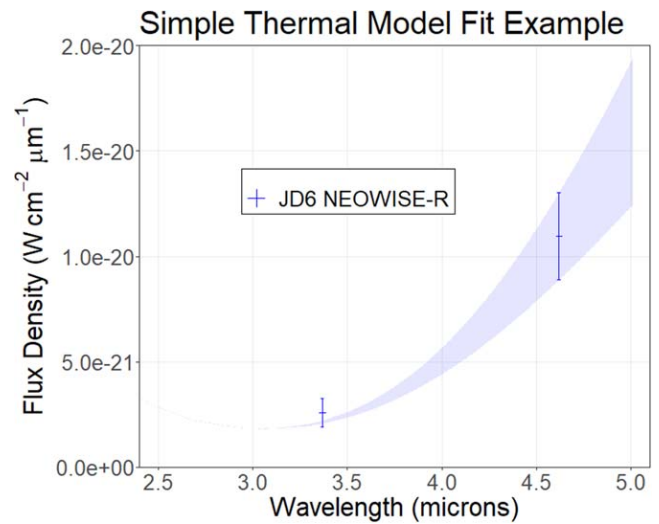

**Figure 4.** Example simple thermal model fit to NEOWISE-R data. Models that fall within the $1\sigma$ uncertainties are considered to fit the data. Here, we show an example of a fully color-corrected NEOWISE-R data set for JD6. The $x$-axis is wavelength in microns, and the $y$-axis is absolute flux density.

We report our model results below in the form of a heatmap for each object at each data processing stage. The $x$-axis is the thermal inertia in TIU, and the $y$-axis for each plot is the visible albedo. Each colored square represents a different model, with the color corresponding to how many NEOWISE data sets can be fit with the given model. Purple squares fit more data sets, and red squares fit fewer data sets. Note that some NEOWISE data sets cannot be fit with any models, and some data sets are fit by models that do not overlap in the parameter space. Therefore, the highest number of data sets fit for any model in the parameter space may be less than the total number of NEOWISE data sets.

#### 4.2. Model Results

We use our simple thermal model to model the NEOWISE data for the NEAs JM8, JD6, and UO1. For JM8, we use an H-magnitude of 15.15 and visible-to-near-IR reflectance ratio of 1.85. Models were run with visible albedos of 0.01–0.10 in increments of 0.01, with thermal inertias of 0–100 TIU in increments of 25 TIU, and with a beaming parameter range of 0.70–1.50. The number of observations and resulting averaged data sets used at each step of the data processing are listed in Table 2. After our initial data pull, 5 NEOWISE-R data sets, comprised of 12 observations, were discarded because they produced obviously erroneous spectral slopes. Of the observations removed after checking the data flags, 6 were removed for violating `cc_flag` rules, 1 was removed for violating `ph_qual` rules, and 1 was removed for violating `qual_frame` rules. Of the observations removed after manually inspecting the images, 13 were removed because another object was overlapping the central object, and 1 observation was removed because it was contaminated with a diffraction spike. An albedo of 0.05 was used for the calculated theoretical blackbody temperature, based on previous modeling results. All model results are shown in Figure 5. Based on radar-derived size measurements and reported H-magnitudes (Table 1), we can further restrict the realistic albedo range to ∼0.01–0.06. This is in agreement with what we find, especially when looking at the full, color-corrected model results, but would rule out some of the higher albedo models.

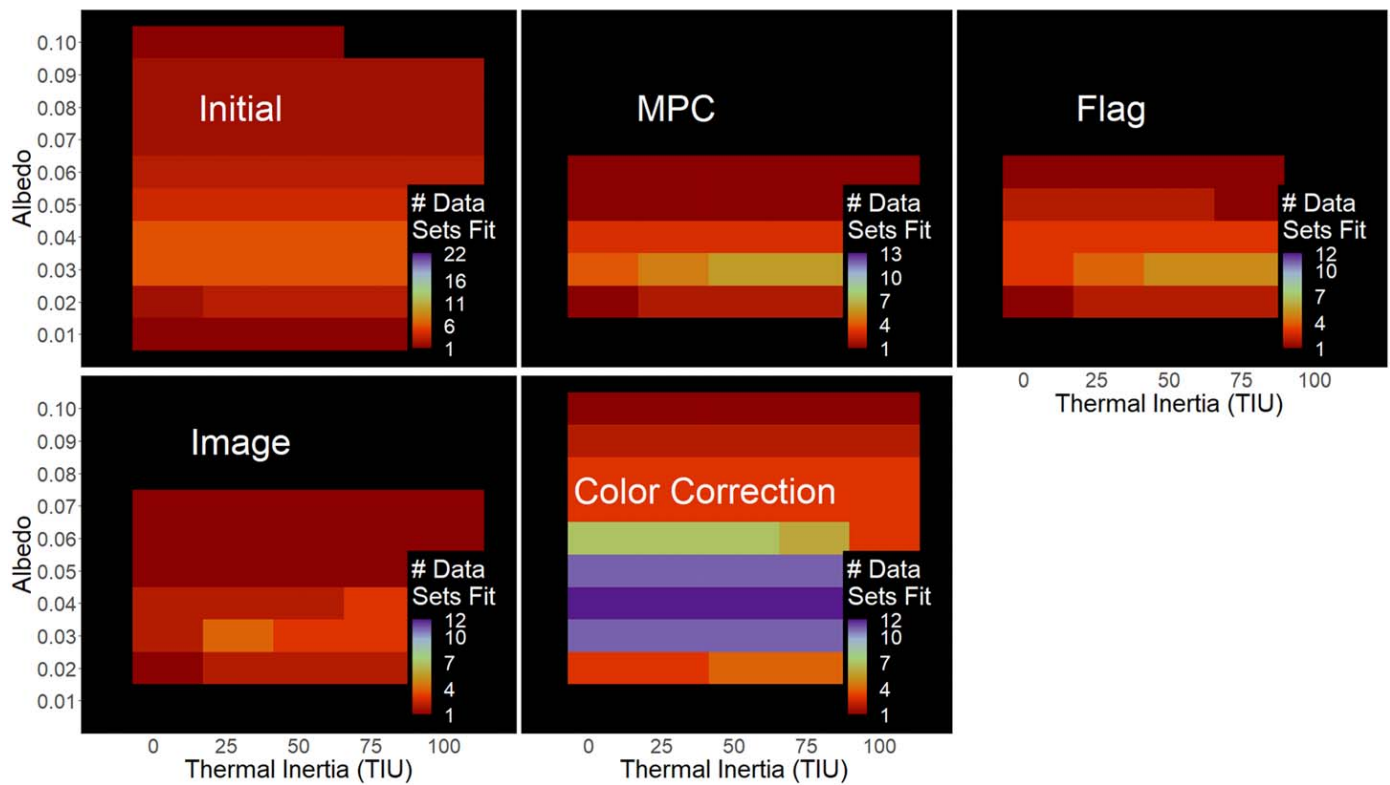

**Figure 5.** Model results for JM8. The *x*-axis is thermal inertia in TIU, and the *y*-axis is visible albedo. Each colored square represents a different model with the color indicating the number of data sets fit by the given model. Note that in some cases the largest number of data sets fit by a model is less than the total number of data sets. Each panel shows model results after a different phase of the data processing.

For JD6, we use an H-magnitude of 17.1 and visible-to-near-IR reflectance ratio of 1.24. Models were run with visible albedos of 0.01–0.31 in increments of 0.02, with thermal inertias of 0–600 TIU in increments of 100 TIU, and with a beaming parameter range of 0.50–1.70. The number of observations and resulting averaged data sets used at each step of the data processing are listed in Table 2. After our initial data pull, 1 NEOWISE-R data set, comprised of 11 observations, was discarded because it produced obviously erroneous spectral slopes. The increase in the number of averaged data sets for the NEOWISE-R data after checking observations against the MPC results from a natural gap opening up in the data after the removal of observations. Of the observations removed after checking the data flags, 4 were removed for violating `cc_flag` rules, 24 were removed for violating `ph_qual` rules, and 15 were removed for violating `qual_frame` rules. Of the observations removed after manually inspecting the images, 4 were removed because another object was overlapping the central object, 2 were removed because the object was not visible in the frame, and 1 observation was removed because there was a streak across the frame. An albedo of 0.18 was used for the calculated theoretical blackbody temperature, based on previous modeling results. All model results are shown in Figure 6. Based on radar-derived size limits and reported H-magnitudes (Table 1), we can further restrict the realistic albedo range to ≳0.04. We note that the albedos most heavily favored by our modeling barely meet this constraint until the full, color-corrected data are used.

For UO1, we use an H-magnitude of 16.7 and visible-to-near-IR reflectance ratio of 1.3. Models were run with visible albedos of 0.05–0.50 in increments of 0.05, with thermal inertias of 0–700 TIU in increments of 100 TIU, and with a beaming parameter range of 0.70–1.20. The number of observations and resulting averaged data sets used at each step of the data processing are listed in Table 2. After our initial data pull, 1 post-cryo data set and 4 NEOWISE-R data sets, comprised of 4 observations and 9 observations respectively, were discarded because they produced obviously erroneous spectral slopes. The observation removed after checking the data flags was removed for violating `cc_flag` rules. An albedo of 0.30 was used for the calculated theoretical blackbody temperature, based on previous modeling results. All model results are shown in Figure 7. Based on radar-derived size limits and reported H-magnitudes (Table 1), we can further restrict the realistic albedo range to $\gtrsim 0.2$, a result in excellent agreement with our full, color-corrected data.

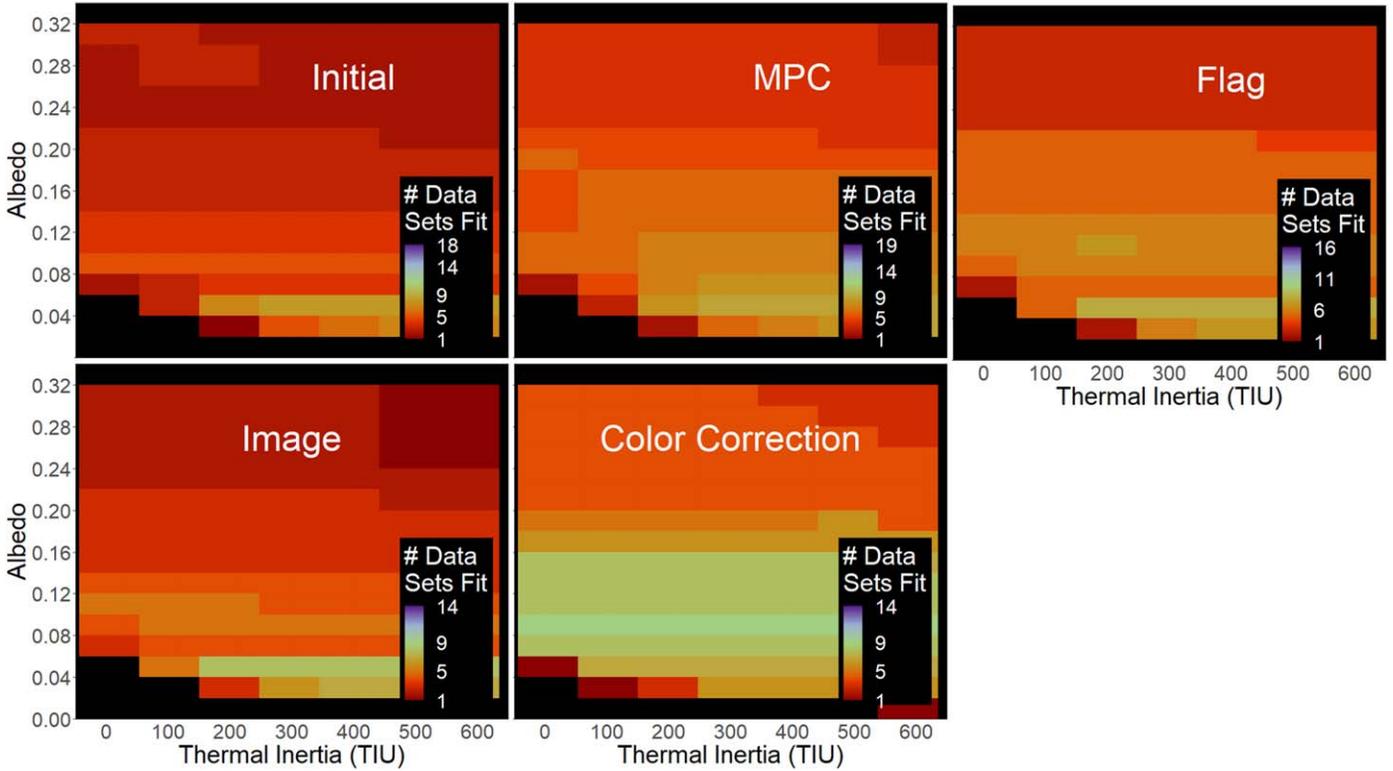

**Figure 6.** Model results for JD6. The *x*-axis is thermal inertia in TIU, and the *y*-axis is visible albedo. Each colored square represents a different model with the color indicating the number of data sets fit by the given model. Note that in some cases the largest number of data sets fit by a model is less than the total number of data sets. This may be due to our simple thermal model being unable to capture aspects of JD6's elongated shape. Each panel shows model results after a different phase of the data processing.

Overall, we highlight that simple thermal model results are affected by failure to follow all of the data processing procedures we have identified as necessary. Primarily, we see that failure to properly vet the data leads to the inclusion of spurious data sets, decreasing model agreement. This is most notably seen for JM8 (Figure 5) where using all initially available data produces 22 data sets, of which only 9 produce overlapping models. However, following all the identified data processing steps reduces the number of data sets to 12, all of which overlap in the fit space. A similar trend is seen for JD6. We also see that overall the shape of the fit space is affected. Namely, we see that by using the full, color-corrected data, the model results are brought into closer alignment with reported albedo (and thus size) values determined via other methods, such as radar.

## 5. Summary

In this work we lay out some of our "lessons learned" from working with NEOWISE data, provide a roadmap for using the data for NEA observations, and provide a novel method for determining NEA color corrections. (Note that all steps, except those relating to the color corrections, are also applicable to MBA observations.) Proper NEOWISE data processing requires a number of steps to ensure that only nonspurious, noncontaminated detections of the actual object are used. These steps are as follows:

1. Check observations against the MPC.
2. Check the automatic pipeline data flags.
3. Check the raw images via visual inspection.
4. Properly average the data, and apply color corrections.

For this last step, we describe an empirical method for determining color correction values that avoids alterations to the model code, additional model complications, and circularity with model results. We do this by incorporating temperature uncertainty into the data uncertainty when determining the color corrections. This is done by comparing blackbody curves fit to IRTF SpeX data of NEAs with those of calculated theoretical blackbody temperatures for these objects. We find that the difference between the two does not exceed 30 K for observations taken at heliocentric distances below 2 au.

Finally, we highlight these results by processing NEOWISE data for three example NEAs: (53319) 1999 JM8, (85989) 1999 JD6, and (137032) 1998 UO1. We step through each of the data processing steps and analyze the remaining data at each step using a simple thermal model. This allows us to see how model results are impacted by each data processing step. Overall, we find that each data processing step is key for eliminating spurious and contaminated detections.

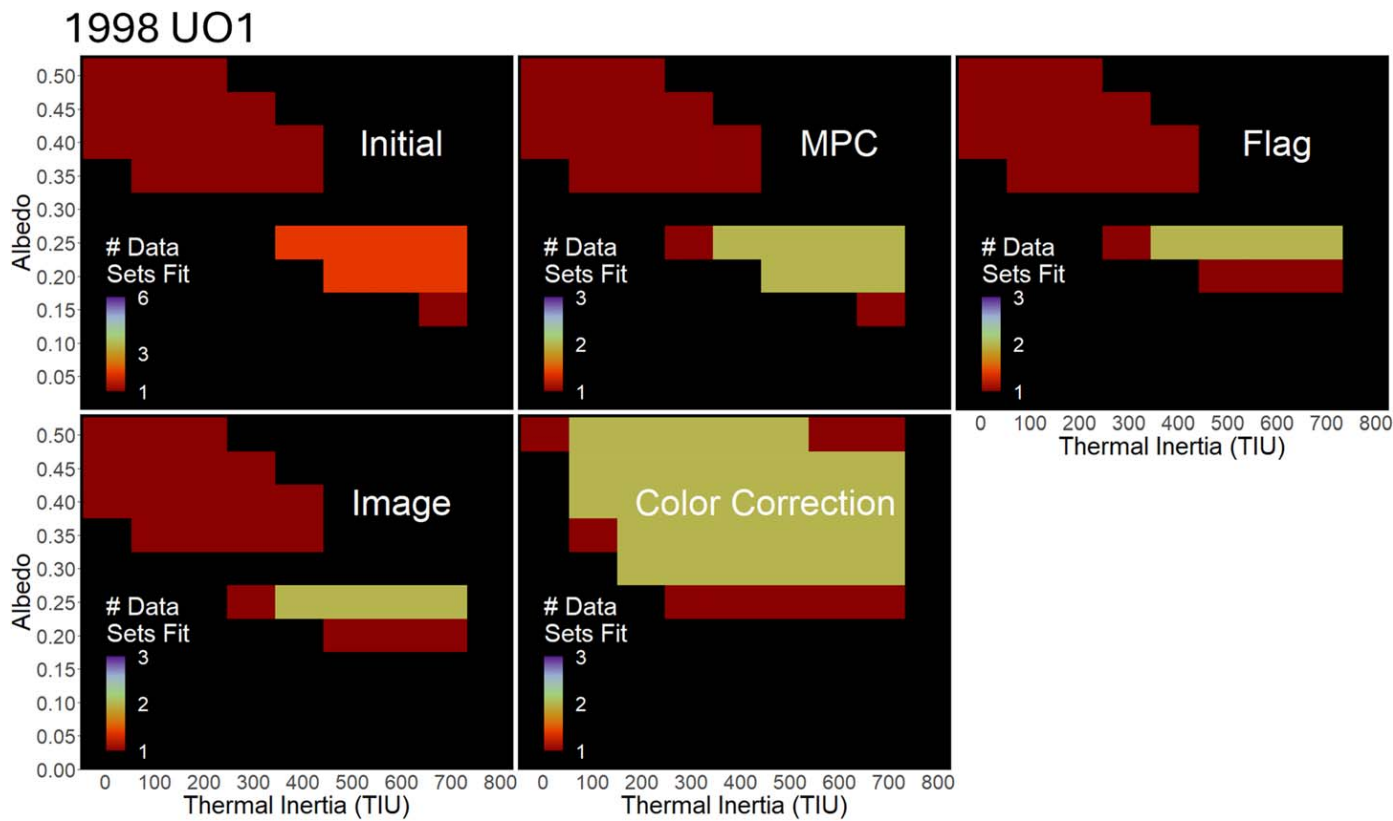

**Figure 7.** Model results for UO1. The *x*-axis is thermal inertia in TIU, and the *y*-axis is visible albedo. Each colored square represents a different model with the color indicating the number of data sets fit by the given model. Note that in some cases the largest number of data sets fit by a model is less than the total number of data sets. Each panel shows model results after a different phase of the data processing.

1. Failure to compare the NEOWISE catalog against the MPC would result in the inclusion of 75 spurious observations (25% of total). Based on our data averaging, this would lead to the inclusion of 11 spurious data sets.
2. Failure to check the data flags provided by the automatic pipeline would result in the inclusion of 52 spurious observations (17% of total). Based on our data averaging, this would lead to the inclusion of four spurious data sets.
3. Failure to visually inspect the raw images would result in the inclusion of 21 spurious observations (7% of total). Based on our data averaging, this would lead to the inclusion of two spurious data sets.
4. Failure to incorporate full color correction changes alters the final averaged data fluxes. As a result, the parameters inferred from use with simple thermal models are altered as well. When full, color-corrected data are used, model results are generally in closer alignment with values reported by other methods and models.

Through this work, we seek to highlight the importance of proper data vetting and data processing for WISE and NEOWISE data when working with NEAs. We also seek to assemble all of this information in a single, easy to locate source, to make it easier for others unfamiliar with the mission architecture to utilize these data in their investigations.

### Acknowledgments

This work was partially funded by the NASA YORPD program (NASA grant 80NSSC21K0658). This material is based upon work supported by the National Science Foundation Graduate Research Fellowship Program under grant No. DGE-2137419. Any opinions, findings, and conclusions or recommendations expressed in this material are those of the author(s) and do not necessarily reflect the views of the National Science Foundation. Authors S.A.M., E.S.H., R.J.V. Jr., Y.R.F., and M.L.H. were Visiting Astronomers at the Infrared Telescope Facility, which is operated by the University of Hawaii under contract 80HQTR19D0030 with the National Aeronautics and Space Administration. S.E.M. was funded by NASA grant 80NSSC23K0377. Thanks to Jenna L. Crowell and Kiana McFadden for additional IRTF observing assistance. Thanks to Joe Masiero for insightful conversation and encouragement about the work presented here.

*Facilities*: NEOWISE, IRTF (SpeX).

*Software*: ggplot2 (https://ggplot2.tidyverse.org/), dplyr (https://dplyr.tidyverse.org/), gWidgets2 (https://github.com/jverzani/gWidgets2), Spextool (M. C. Cushing et al. 2004).

### ORCID iDs

Samuel A. Myers ● https://orcid.org/0000-0001-8500-6601
Ellen S. Howell ● https://orcid.org/0000-0002-7683-5843
Yanga R. Fernández ● https://orcid.org/0000-0003-1156-9721
Sean E. Marshall ● https://orcid.org/0000-0002-8144-7570
Christopher Magri ● https://orcid.org/0000-0002-2200-4622
Ronald J. Vervack, Jr. ● https://orcid.org/0000-0002-8227-9564
Mary L. Hinkle ● https://orcid.org/0000-0003-0713-2252